%% file: demandforllm.tex
\documentclass[12pt]{article}
\usepackage{amsmath}    
\usepackage{amssymb}    
\usepackage{natbib}

\setcitestyle{aysep={}}

\usepackage{booktabs}
\usepackage{siunitx}
\usepackage{float}
\usepackage{fullpage}
\usepackage{tabularray}
\usepackage{siunitx}  
\usepackage[normalem]{ulem}
\usepackage{longtable,caption}
\usepackage{cleveref}
\usepackage{graphicx}   
\usepackage{subfig}     
\newcommand{\smalllongtable}{%
  \small               
  \setlength{\LTpre}{0pt}\setlength{\LTpost}{0pt}}

\NewTableCommand{\tinytableDefineColor}[3]{\definecolor{#1}{#2}{#3}}
\newcolumntype{d}{S[
    input-open-uncertainty=,
    input-close-uncertainty=,
    parse-numbers = false,
    table-align-text-pre=false,
    table-align-text-post=false
 ]}

\begin{document}
\title{Demand for LLMs: Descriptive Evidence on Substitution, Market Expansion, and Multi-Homing\\
(Preliminary, comments very much welcome.)}
\author{Andrey Fradkin\thanks{afradkin@gmail.com, Boston University and MIT IDE. Thanks to Tom Cunningham, Dean Eckles, and Daniel Rock for initial feedback on this work. LLMs were very helpful in conducting this analysis.}}
\date{\today{}}                       
\begin{titlepage}\maketitle

\begin{abstract}
    This paper documents three stylized facts about the demand for Large Language Models (LLMs) using data from OpenRouter, a prominent LLM marketplace. First, new models experience rapid initial adoption that stabilizes within weeks. Second, model releases differ substantially in whether they primarily attract new users or substitute demand from competing models. Third, multi-homing—using multiple models simultaneously—is common among apps. These findings suggest significant horizontal and vertical differentiation in the LLM market, implying opportunities for providers to maintain demand and pricing power despite rapid technological advances.
\end{abstract}

\end{titlepage}
\section{Introduction}
Massive investments have been undertaken by companies such as Anthropic, Google, and OpenAI to advance the state of the art in artificial intelligence. While few doubt that these investments will yield better AI models, whether these models will generate extraordinary profits remains uncertain. In one state of the world, AI models are undifferentiated and competition will result in a race to the bottom, with prices equal to marginal costs. In another state of the world, AI models are heterogeneous and customers have differentiated preferences for one model provider versus another. Which of these worlds we live in is an empirical question. 

In this paper, I present descriptive evidence about demand for Large Language Models (LLMs). I use the release of three models (Claude Sonnet 3.7, Gemini Flash 2.0, and Gemini 2.5 Pro) to document three stylized facts about the demand for AI models. 
My analysis suggests that, at least at present, LLMs are both horizontally and vertically differentiated from each other. 

The stylized facts are listed below.
\begin{enumerate}
\item Improved models are adopted quickly, with increased demand stabilizing within a few weeks.
\item New model releases differ in the degree to which they cause substitution from existing models or expand the market.
\item There is substantial multi-homing, with users of the same app employing a mix of models. 
\end{enumerate}

To conduct this analysis, I use data scraped from OpenRouter, which is a marketplace for LLMs used by over a million users either directly or through popular apps such as Cline and Roo Code. OpenRouter provides an API that helps app developers and other users manage their interactions with a variety of LLMs, including providing tools for routing API calls across models and providers depending on price and latency. My dataset consists of daily usage by model between January 11, 2025 and April 11, 2025. In addition to daily usage, I observe the dates when new models are launched, and for a select set of applications and models, their model specific weekly demand. 

The analysis in this paper is descriptive. I zoom in on several model release events that happened in 2025, and document demand patterns around these release events. Specifically, I focus on the releases of Claude 3.7 Sonnet, Gemini 2.0 Flash, and Gemini 2.5 Pro. For the analysis of multi-homing, I focus on two popular coding apps (Cline and Roo Code) and two popular chat apps (SillyTavern and Shapes Inc). 

The facts documented above highlight a key tension in the demand for LLMs --- how to reconcile persistent demand with seemingly low switching costs. For example, model providers such as Anthropic have achieved some level of demand stickiness despite not being cheap (Sonnet 3.7 is priced at \$3 per million tokens (MTok) while Gemini 2.5 Pro is priced at \$1.25 (MTok)) and not being at the top of benchmarks (for example Claude 3.7 (without thinking) scores 60.4\% on the Aider polygot coding benchmark while Gemini 2.5 Pro scores 72.9\%). A key question for competition is whether persistent demand reflects important differences in quality not measured by benchmarks. For example, maybe Claude has better `vibes' than Gemini for coding. Alternatively, there could be other behavioral factors that cause model use persistence. For example, persistent demand may reflect branding or superior integration of the model with specific coding tools.

While this analysis is a first look at LLM demand, it does have many limitations. First, OpenRouter does not have data on consumer usage of LLMs through native apps such as ChatGPT. Second, some popular apps such as Cursor do not use OpenRouter (or at least do not publicly disclose their usage). As a result, my analysis is only looking at part of the market. Finally, OpenRouter API calls come disproportionately from apps used for programming or character personas, and as a result are not representative of LLM demand for other use cases.

My work contributes to ongoing work in the economics of AI. \citet{eloundou2024gpts} and \citet{handa2025economic} consider the potential task level exposure to LLMs, while \citet{acemoglu2024simple} embeds AI into a task based macroeconomic model. Several papers have considered how AI can affect productivity for a variety of tasks (\citet{brynjolfsson2025generative}, \citet{dellacqua2023navigating}, and \citet{handa2025economic}). My work complements this work by considering the demand for better AI models. 

Separately, a large literature in economics and marketing considers modeling demand (e.g., \citet{dube2019microeconometric}). Of particular importance in demand modeling is the diversion ratio (\citet{conlon2021empirical}), which is a measure of how demand shifts with changes to the availability or prices of products. I provide some of the first evidence of diversion ratios in the market for LLMs.

\section{Institutional Background and Data Preparation}

In this section, I first describe the data scraping and cleaning procedure used for the main dataset. I then describe the cleaned dataset.

OpenRouter is a private company that describes itself as ``the unified interface for LLMs,'' with a subtitle of ``Better prices, better uptime, no subscription.'' What this means in practice is that OpenRouter provides a standardized API for calling any of hundreds of models. This has a variety of advantages for developers in addition to simplicity. One advantage is that for models offered by several providers (e.g., gpt4o on Azure vs OpenAI), OpenRouter can smartly route the API call based on latency, price, or throughput. OpenRouter also gives developers options to use some models as fallback for others under specified circumstances.

On its site, OpenRouter provides a set of rankings that specify the top models by tokens used over time. Of particular interest is that the webpage for each model contains data on daily tokens used starting from January 11, 2025. The model page also includes other pertinent information such as price, uptime statistics, and top apps using the model per week (including tokens used). My main dataset comes from daily scrapes of these model pages, where there are 296 models in total.

From this dataset, I remove old models (those created before Jan 1, 2024) and I merge beta and non-beta versions of the same model. This leaves me at 249 models and 16,584 model day observations. \Cref{tab:summary_stats} presents summary statistics at the model-day level. There is enormous variation in the demand for models, with some models not being used at all, while other models having prompt tokens on the order of 55 MTok per day (Anthropic's Claude 3.7 Sonnet has the highest average prompt token usage in the sample, while Google's Gemini 2.0 Flash 001 has the highest average completion tokens). The number of tokens used for prompts is generally higher than the number of tokens used for completion. Context windows also vary quite a bit, with Google models generally having the largest context windows. Lastly, pricing is highly variable. OpenRouter hosts models that are free, but also hosts some of the most expensive models including Claude Sonnet 3.7 and OpenAI o1-pro.

\begin{table}[htbp] 
    \centering 
    \caption{Summary Statistics by Model-Day} 
    \label{tab:summary_stats} 
    \input{tables/summary_stats_overall.tex}
    
    \smallskip
    \small{\emph{Notes:} Each observation is a model by day. MTok stands for million tokens.}
\end{table}

\Cref{tab:summary_statsclass} and \Cref{tab:summary_statsprovider} display these summary statistics broken out by model class (state-of-the-art (SOTA), fast \& cheap, and old), and provider.\footnotemark{} On average, Google models are much cheaper than Anthropic or OpenAI models, with other models being the cheapest. It is important to note that many of the other models are open source, smaller, and are served by a variety of providers such as Groq, Lambda, and DeepInfra.

\footnotetext{There isn't an obvious way to classify models, since their relative status changes over time. I defined state of the art as models that were at or close to the frontier of benchmarks at any point during the sample period. \Cref{tab:allmodels} displays all the models in the sample and my categorization.}

Lastly, for data about specific applications and the models they use, I leverage the fact that OpenRouter reports the weekly top (public) apps using each model,\footnotemark{} as well as the tokens used. By taking the union of these observations across models, I am able to compile app-specific model usage. This procedure isn't perfect, since I won't be able to see demand when an app is not in the top 20 for a particular model. That said, since the platform is dominated by a few models, and a few apps, I am able to obtain a detailed demand profile for top apps.

\footnotetext{Not all apps allow OpenRouter to publicly display their usage data.}

The top apps in my sample, along with their token usage are listed in \Cref{tab:topapps}. Three of the most used apps are for coding (Cline, Roo Code, and liteLLM). There are four apps that are useful for chat and/or personals (Shapes Inc, SillyTavern, Cub AI, and Chatroom). DocsLoop is an app used for data extraction from documents, Fraction AI is a platform for AI agents to compete, and Fish Audio is an audio AI provider.

\begin{table}[!htbp] 
    \centering 
    \caption{Top Apps} 
    \label{tab:topapps} 
    \begin{minipage}{\linewidth}
        \centering
        \input{tables/top_10_apps_simple.tex}
        
        \smallskip
        \small{\emph{Notes:} Usage calculated for the models in which the app is in the top 20 public apps. This usage data comes from the week prior to April 11, 2025.}
    \end{minipage}
\end{table}

\begin{figure}[htbp]
    \centering
    \subfloat[Token Usage]{\includegraphics[width=0.48\textwidth]{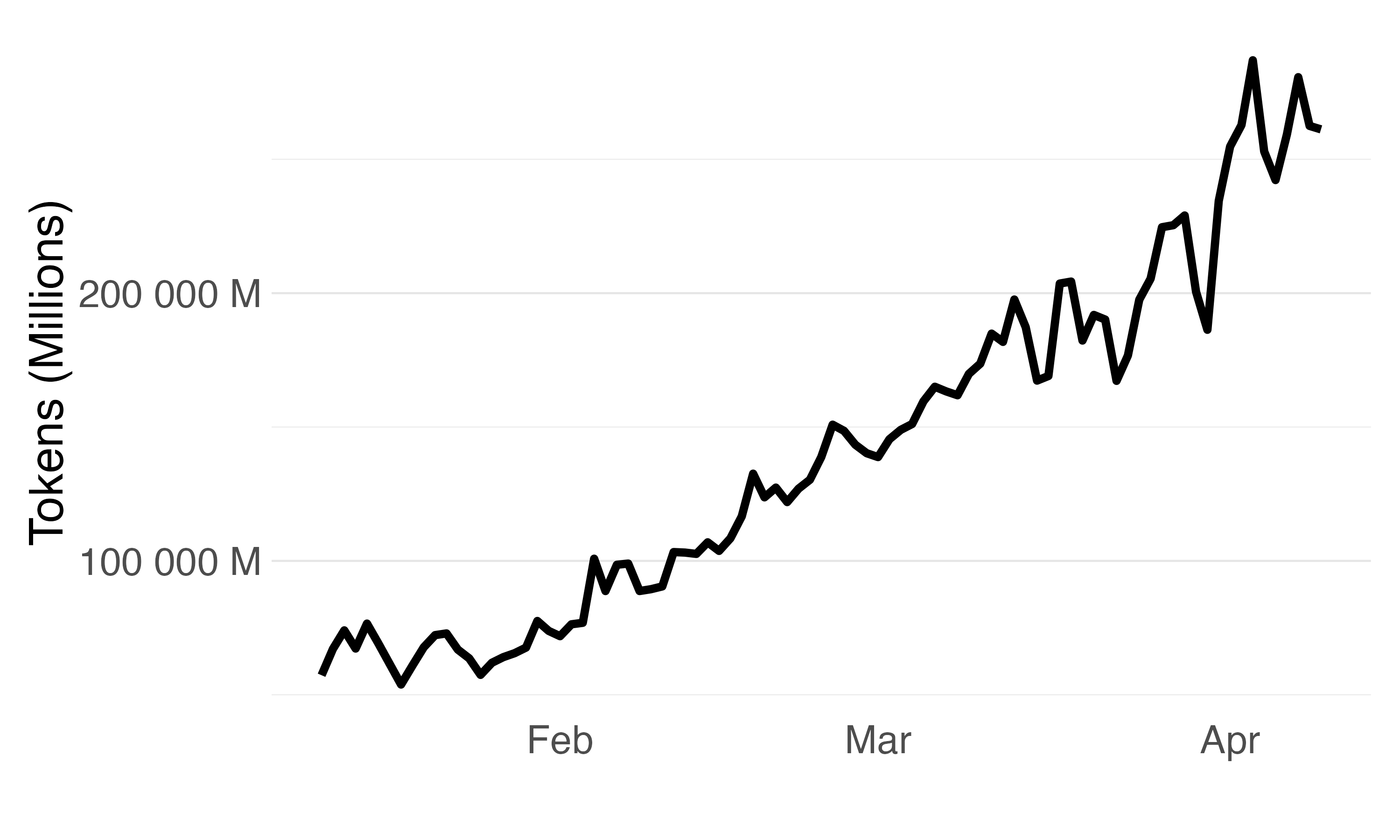}}
    \hfill
    \subfloat[Revenue]{\includegraphics[width=0.48\textwidth]{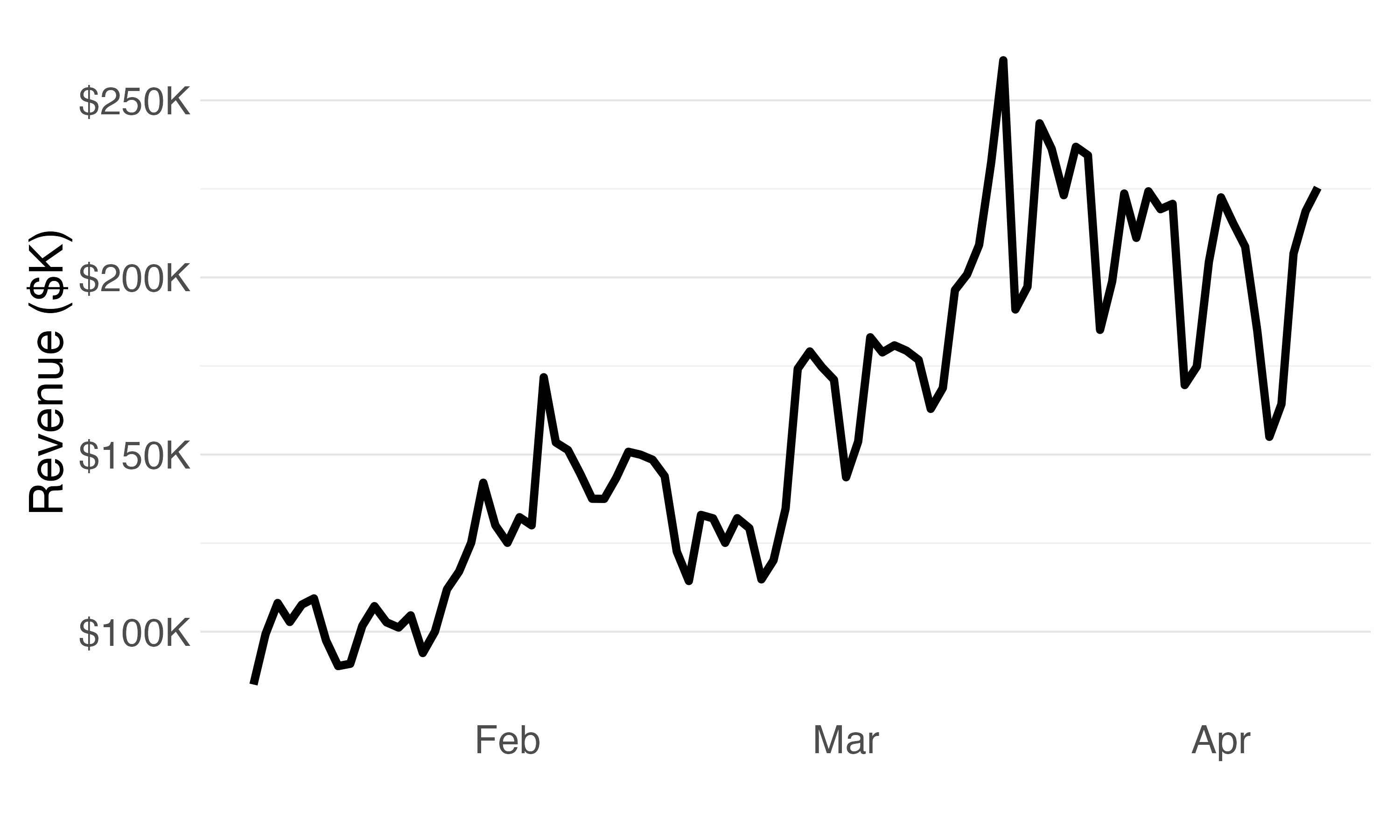}}
    \\ 
    \subfloat[Top Models vs Other]{\includegraphics[width=0.8\textwidth]{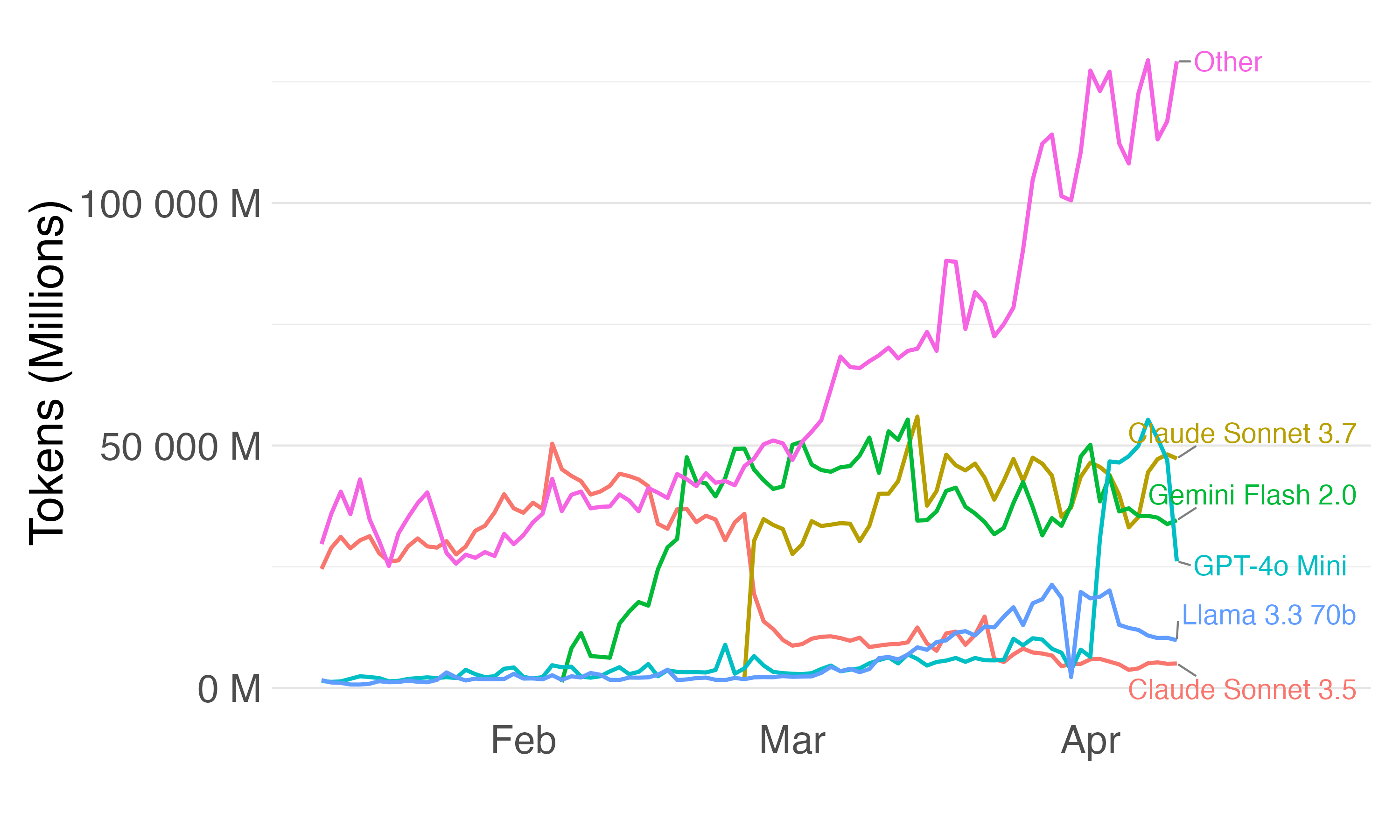}} 
    \caption{OpenRouter Growth (Jan-Apr 2025)}
    \label{fig:platform_growth}
\end{figure}

Lastly, it's helpful to consider the total usage of the platform and how it's evolving over time. The platform has shown consistent growth throughout the analysis period, as evidenced in Figure \ref{fig:platform_growth}. Total revenue\footnotemark{} increased from approximately \$85K to over \$200K per day between January and April 2025, though with significant volatility. Similarly, token usage has grown steadily from around 50 billion to over 250 billion tokens per day. This growth reflects both increased adoption of existing models and the introduction of new, more capable models during the period.

\footnotetext{Total revenue is calculated by using the price as of April 2025 for input and output tokens and multiplying it by the corresponding number of tokens. Reasoning tokens are not included in this calculation.}

\section{Case studies and stylized facts about LLM Demand}

The bulk of the analysis consists of case studies, namely the demand patterns of models around the release of other models. I believe that this sort of analysis is informative since new model releases get a lot of media attention, offer new capabilities, and cause large shifts in demand. 

Nonetheless, it is useful to state the limitations of case study analysis. Formally, this interrupted time-series analysis requires many assumptions. Two I'd like to state explicitly are the following. First, there should be no concurrent events which occurred in the market in the same time period as the release of a model, which could have affected demand to a similar degree. This assumption might be violated if, for example, a large app started using OpenRouter in the same time period. Second, the release of a new model should not cause people to switch to using OpenRouter or away from using OpenRouter. These two assumptions need to be evaluated separately for each case study.

\subsection{Case Studies}

\begin{figure}[!htbp]
    \centering
    \includegraphics[width=.8\textwidth]{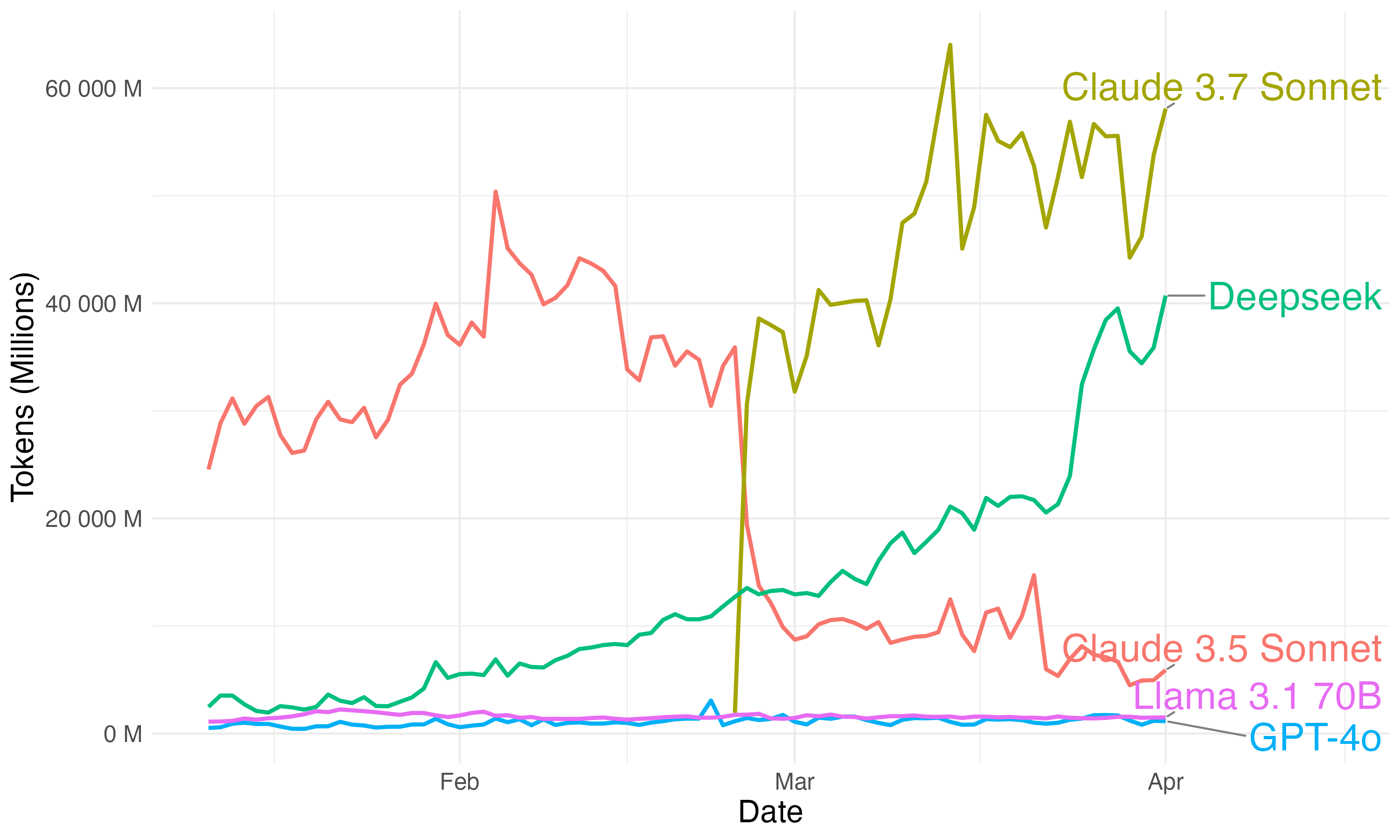}
    \caption{Token Usage for a Models Following Claude 3.7 Sonnet Release (Feb. 2025)}
    \label{fig:claude_case_study}
\end{figure}

\textit{Claude 3.7 Release:}
Claude 3.7 is a frontier model that was released by Anthropic on Feb 24, 2025. The model was branded as having particular strengths in coding and front-end web development. Figure \ref{fig:claude_case_study} shows the impact of Claude 3.7 Sonnet's release on the token usage patterns of leading models (\Cref{fig:claude_case_study_log} displays the results on a log scale). As a comparison, it plots demand for several other advanced models: Claude 3.5, Anthropic's previous model, OpenAI's model GPT-4o and Deepseek.

\textit{Gemini 2.0 Flash Release:}
Gemini 2.0 Flash was fully released by Google on February 05, 2025, in concurrence with the release of Gemini 2.0 Flash-Lite which is a more cost-efficient variant. Preview versions of this model were available to a limited extent on OpenRouter prior to this date. Gemini 2.0 is considered a strong and quick model for its price, even though it does not achieve state-of-the-art performance on capabilities benchmarks. 

\begin{figure}[!htbp]
    \centering
    \subfloat[Token Usage for Models Following Gemini 2.0 Flash Release]{\includegraphics[width=.8\textwidth]{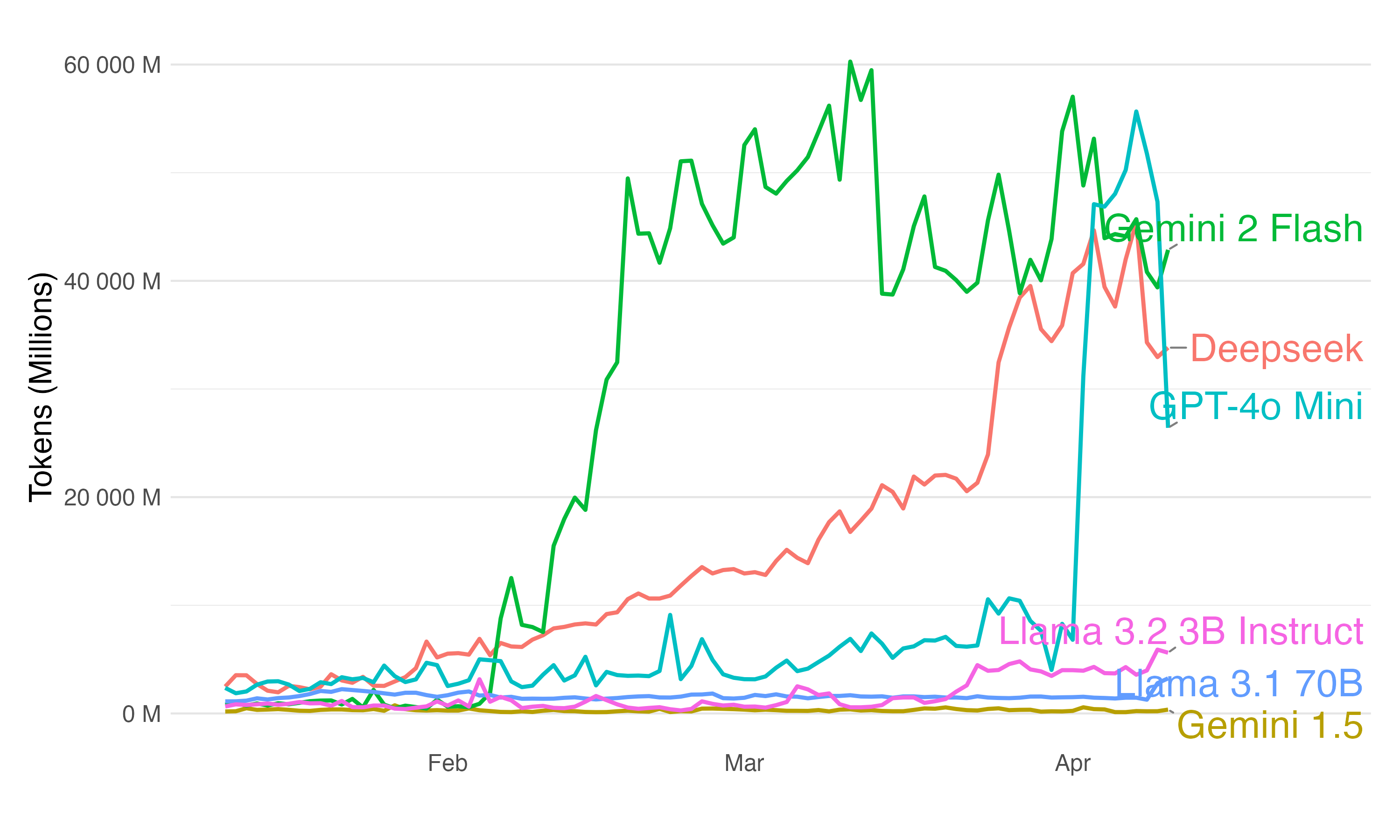}    \label{fig:gemini_20a_case_study}
    }
    
    \subfloat[Disaggregated view of Gemini 2.0 Flash models]{\includegraphics[width=.8\textwidth]{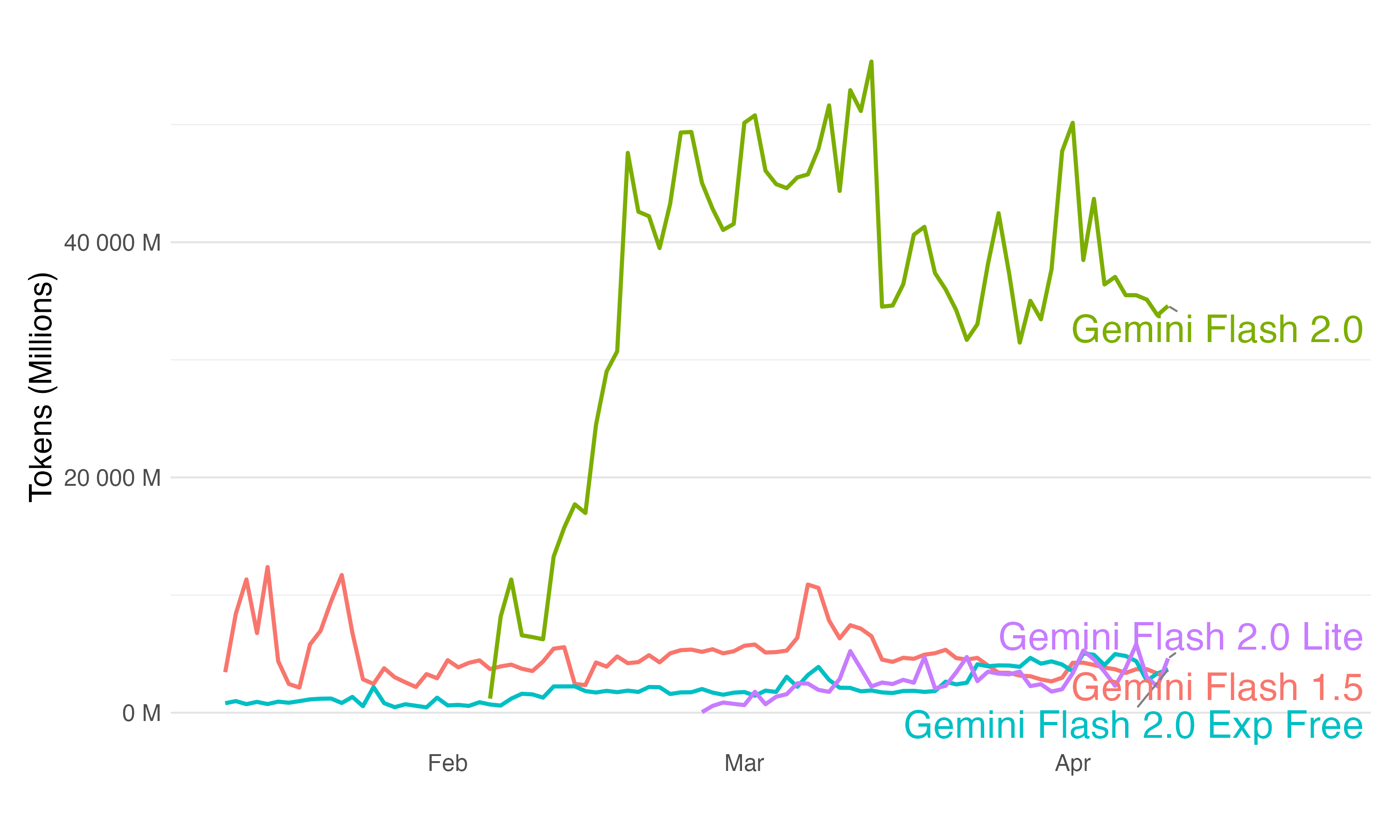}    \label{fig:gemini_20b_case_study}
    }
    \caption{Gemini 2.0 Flash Release (Feb. 2025)}
\end{figure}

Figure \ref{fig:gemini_20a_case_study} shows the demand for Gemini 2.0 Flash and other potentially comparable models including Deepseek, Gemini 1.5, GPT-4o mini and two Llama variants (\Cref{fig:gemini_flash_case_study_log} displays the results on a log scale). Note, I aggregate three related Gemini 2.0 Flash models --- Flash, Flash Experimental Free, and Flash Lite. We see a rapid rise in Gemini 2.0 Flash demand (green line) after its release. To help interpret the increase in demand, I plot the demand for related models separately (\Cref{fig:gemini_20b_case_study}). The increase is driven by the main Gemini Flash 2.0 model and not by the variants.

\textit{Gemini 2.5 Pro Release:} Gemini 2.5 Pro is Google DeepMind's first ``thinking'' Gemini model.\footnotemark{} An experimental build shipped to Google AI Studio and Vertex AI on March 25 2025. OpenRouter began offering the free experimental variant a few days after the initial rollout, giving developers access to the model. This experimental version is rate limited for developers, meaning that they couldn't use it fully. Gemini Pro 2.5 Preview was released later without these rate limits. 

\footnotetext{Thinking or reasoning models are trained to use output tokens to consider their answer prior to answering a prompt.}

\begin{figure}[!htbp]
\centering
\includegraphics[width=.8\textwidth]{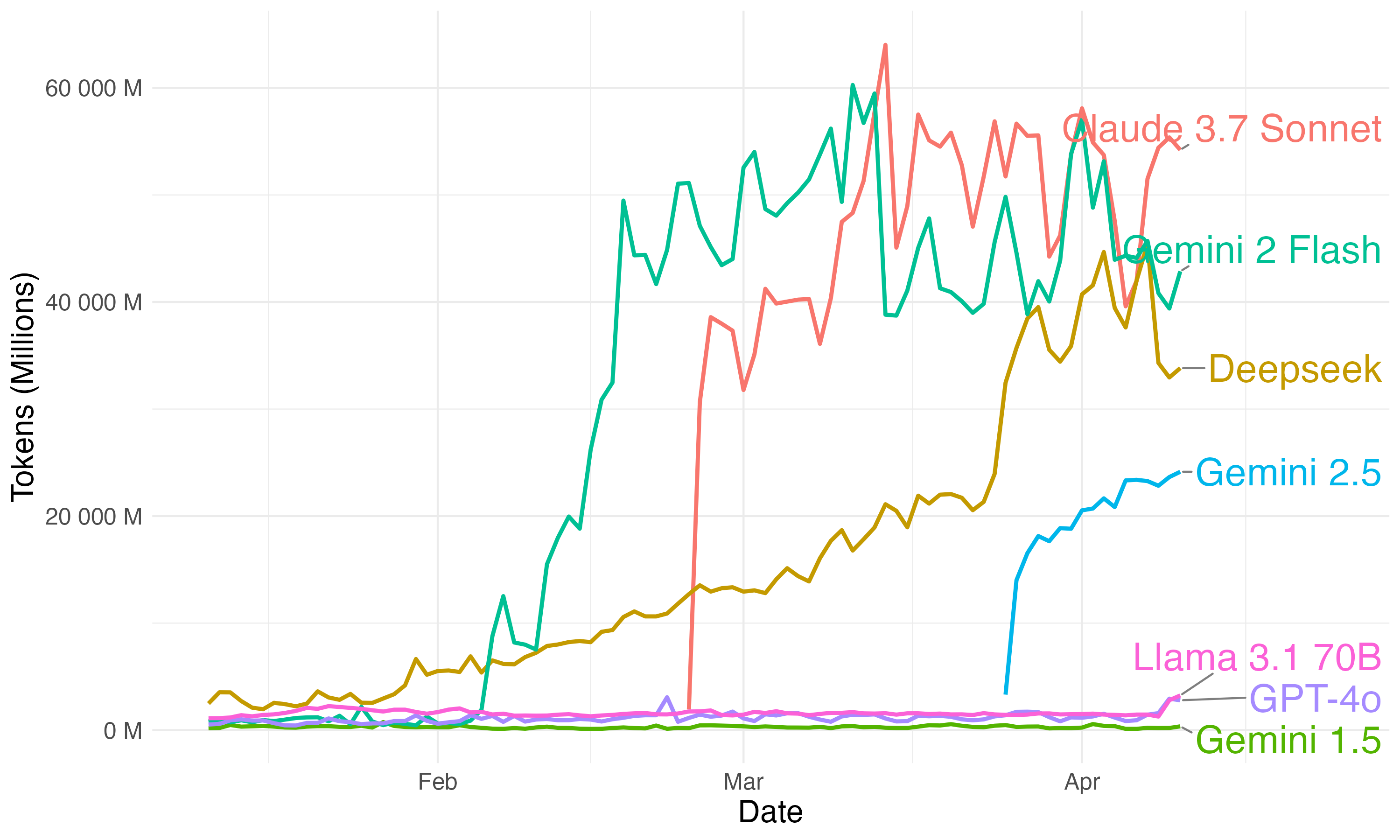}
\caption{Token Usage for Models Following Gemini 2.5 Flash Release (March 2025)}
\label{fig:gemini25_case_study}
\end{figure}

Figure \ref{fig:gemini25_case_study} shows demand for Gemini 2.5 (teal) alongside other comparable models including Claude 3.7 Sonnet, Gemini 2.0 Flash, Deepseek, and GPT-4o (\Cref{fig:gemini25_case_study_log} displays the results on a log scale). Demand for Gemini 2.5 Pro rises quickly, but remains below that of Calude 3.7 Sonnet, Deepseek, and Gemini 2.0 Flash.

\subsection{Three facts about LLM Demand}
\paragraph{Fact 1: The release of better models results in rapid adoption.}

Across all three case studies, adoption of the new model occurs quickly, with an initial jump in log tokens that stabilizes within a few weeks of a model's release on OpenRouter. This is true even when the model provider has not previously been known for frontier models. For example, Google's Gemini 2.5 was the first of its model to achieve the state-of-the-art performance on a variety of LLM benchmarks. Yet even though many may not have expected it to garner fast adoption, it did so relatively quickly. This fast adoption curve is also evident for fast and cheap models, such as Gemini 2.0 Flash.

Note that while the initial jump is rapid, growth in model usage can continue afterward. For example, usage of Gemini 2.5 continues to increase after the initial jump in demand, but at a slower rate. This continued growth of Gemini 2.5 is consistent with the overall growth of the OpenRouter platform, as well as with ongoing learning and optimization by users of the LLMs. 

\paragraph{Fact 2: New model releases differ in the degree to which they cause substitution from existing models or expand the market.}

For each of the three case studies, I picked comparison models that would ex-ante potentially be strong competitors. If models are substitutes, we would expect to see that demand for these models falls as the new model is adopted. We see quite different trends across the three releases, with some models expanding the market and other models creating obvious substitution.

First, consider the release of Claude 3.7 Sonnet (\Cref{fig:claude_case_study}). We see an obvious decline in the usage of Claude 3.5 Sonnet that is concurrent with the adoption of 3.7. Yet, we see no obvious movement in potentially competitive models such as DeepSeek, GPT-4o, and Llama 3.1 70B. My interpretation of this is that substitution for Claude comes mainly within model class (Claude Sonnet), and not from other models. This within Claude Sonnet substitution may be due to a variety of factors. For example, the set of users who are willing to pay for the best coding LLM may have been primarily using Sonnet 3.5 and naturally found it easy to switch to Sonnet 3.7. Alternatively, there may be an Anthropic or Sonnet brand effect that is strong for a subset of users.

Second, consider the release of Gemini 2.0 Flash (\Cref{fig:gemini_20a_case_study}). This figure includes several comparison models, including Gemini Flash 1.5 and Llama, that might ostensibly be considered either cheap or fast, just like Gemini 2.0 Flash. Yet none of these models experience a drop in demand that is concurrent with the rise of Gemini 2.0 Flash. In fact, demand for Gemini 2.0 quickly rises to be greater than the demand for all the other models. One way to interpret this is that Gemini 2.0 greatly expanded the market for LLMs.

Lastly, consider the release of Gemini 2.5 (\Cref{fig:gemini25_case_study}. I observe rapid adoption of the model, but that demand for other models did not seem to change in an obvious way concurrent with the release. This is once again evidence for a market expansion effect of certain LLMs. Of course, since this version of Gemini 2.5 was free and rate limited, we will have to wait to get more data on the full release to learn more about the long-run effects of Gemini 2.5's release.

\paragraph{Fact 3: There is substantial multi-homing. The same app uses a mix of available models.}

The final fact I'd like to document is that there is substantial multi-homing in LLM use. In particular, the same app typically uses a mix of models. To see this, I focus on two popular programming apps (Cline and Roo Code) and two popular chat and persona apps (Shapes Inc and SillyTavern).

\begin{figure}[!htbp]
    \centering
    \includegraphics[width=.8\textwidth]{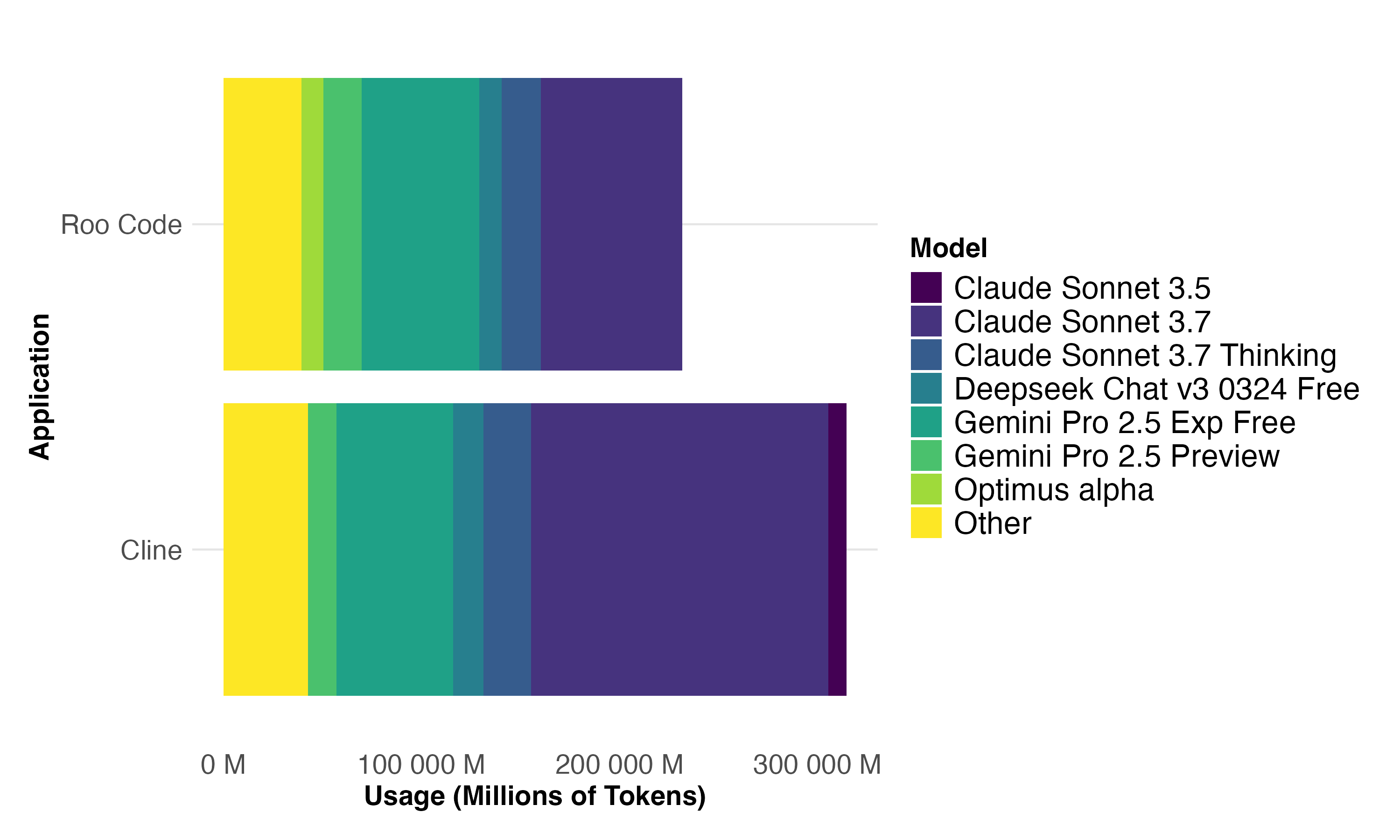}
    \caption{Tokens Used by Model and App (Coding)}
    \label{fig:codingcasestudy}
\end{figure}

\Cref{fig:codingcasestudy} shows the tokens used by model for Cline and Roo Code. Both coding apps have a mix of models used. Claude Sonnet 3.7 is the most popular for both apps, but Gemini Pro 2.5 is a strong second contender. In addition, Deepseek, Claude Sonnet 3.5, and Optimus Alpha (an early version of Gpt4.1) all saw substantial usage. This data shows that users don't perceive just one model to be the best for coding.

\begin{figure}[!htbp]
    \centering
    \includegraphics[width=.8\textwidth]{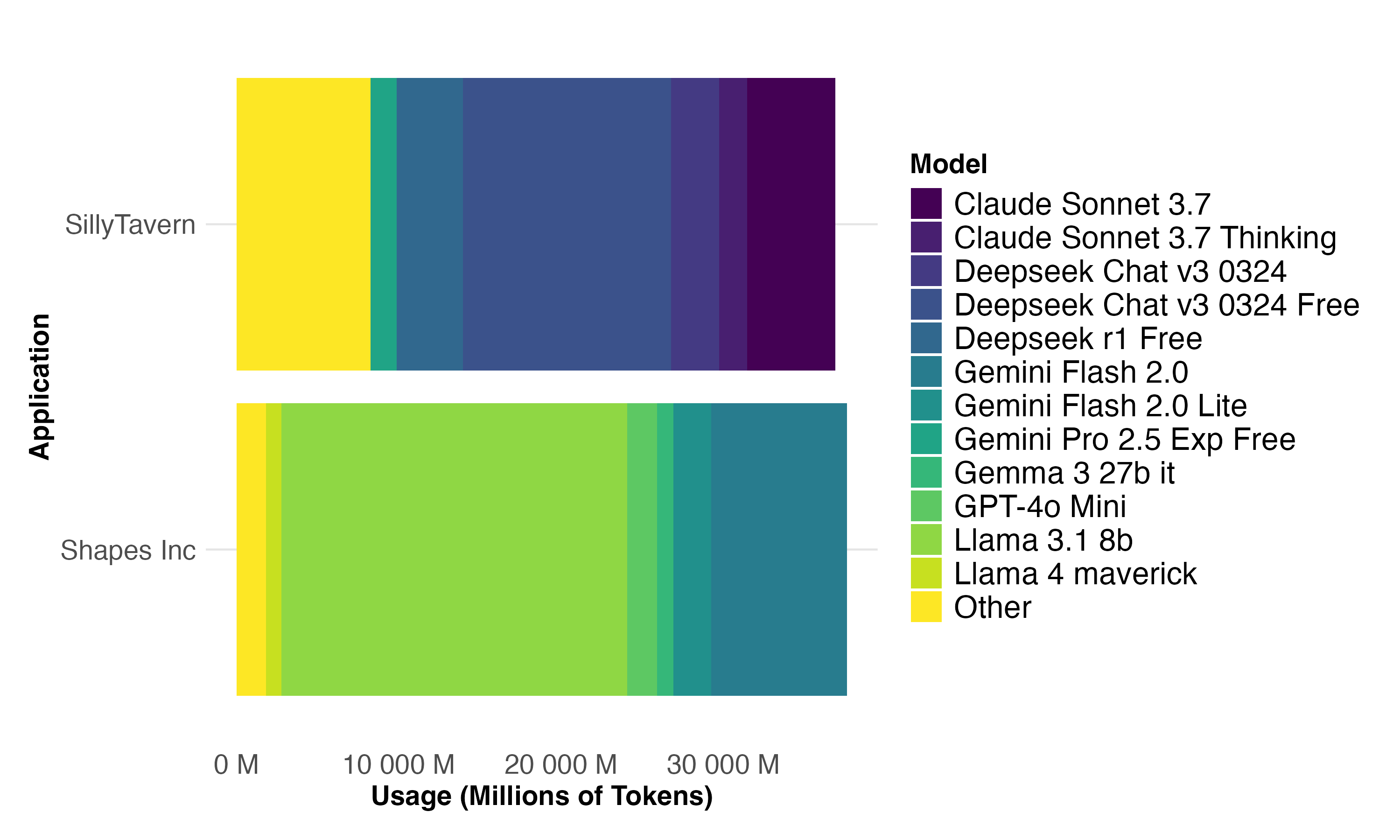}
    \caption{Tokens Used by Model and App (Chat and Personas)}
    \label{fig:chatcasestudy}
\end{figure}

\Cref{fig:chatcasestudy} shows similar multi-homing for SillyTavern and Shapes Incs. Both apps have a variety of models used. SillyTavern has the greatest usage for DeepSeek and Claude, with a substantial share of other models in the long tail (yellow bar). Shapes Inc has quite a different pattern with a max of mainly Gemini Flash 2.0 and Llama 3.1 8b. The overall pattern for Shapes Inc suggests a preference for cheap and fast models. Even within these models, however, there is a lot of variety with a concentration in Llama and Gemini 2.0 Flash.

To summarize, there is a large degree of multi-homing in four popular apps using LLMs. Multi-homing is prevalent, and each app tends to use different combinations of LLMs. 

\section{Implications for competition in AI and open questions.}

It is useful to think of competition in LLMs through the lens of industrial organization economics. In a standard static Bertrand-Nash pricing equilibrium, firms are able to earn profits when their product is differentiated from other products or when some firms have cost advantages over other firms. In my analysis I've focused on the demand side of the market, ignoring the cost side.

In markets where products are differentiated from each other, higher priced products are able to garner substantial demand. We see that this is indeed the case, with Claude Sonnet 3.7 being relatively expensive and having high demand. Differentiation is typically modeled in a vertical dimension (there is one dimension of quality which everyone agrees upon) and a horizontal dimension (in which some people prefer one model while others prefer another at the same price). 

For the LLM market, the existence of horizontal differentiation is particularly important. The reason for this is that smarter and more capable models are released many times a year, and no one provider can hope to always offer the most capable model. In contrast, if models are horizontally differentiated, then providers who don't offer the most capable model can still garner substantial demand with a price markup. 

The three stylized facts suggest that demand for LLMs is differentiated horizontally in addition to vertically. New models are adopted quickly, but substitution patterns differ greatly across models. Sonnet models seem to be substitutes for each other but not for other models. In contrast, Gemini Flash 2.0 and Pro 2.5 expand the market for LLMs without any obvious cannibalization effects. If the most capable models at a given price point received all the demand, we would see a lot more substitution and one model dominating every price point. Additionally, the presence of LLM multi-homing across four popular apps suggests that different LLMs have different use cases for serving end-users. 

This work is meant to offer an initial look at LLM demand, but many open questions remain. The first, and perhaps the most obvious, question is how demand for LLMs responds to price. In order to learn about this, we would need to observe price changes within a model. Even then, there may be differences between short-run price elasticities (responses of demand to short-run pricing changes) and long-run price elasticities (responses of demand to sustained price reductions as computing capacity increases). 

A second question concerns the dimensions of horizontal differentiation in the LLM market. Some obvious ones include speed, context window, down-time, throughput, and license terms. Others include more subjective perceptions such as `vibes' and the presence of app-model integration (when for example a coding tool is optimized to use one LLM and not another). Relatedly, capacity constraints may play an important role in this market, and may limit the extent to which one model provider can capture the market.

Lastly, OpenRouter is just one marketplace for LLMs, and does not see the entire market. It remains an open question whether users that directly interface with model providers such as Anthropic, Google, and OpenAI exhibit similar demand patterns. 

\bibliography{demandforllm}

\appendix
\counterwithin{table}{section}
\counterwithin{figure}{section}
\renewcommand{\thetable}{A\arabic{table}}
\renewcommand{\thefigure}{A\arabic{figure}}
\section{Addition Figures}
\subsection{Claude 3.7 Sonnet Release}

\begin{figure}[H] 
    \centering
    \includegraphics[width=0.8\textwidth]{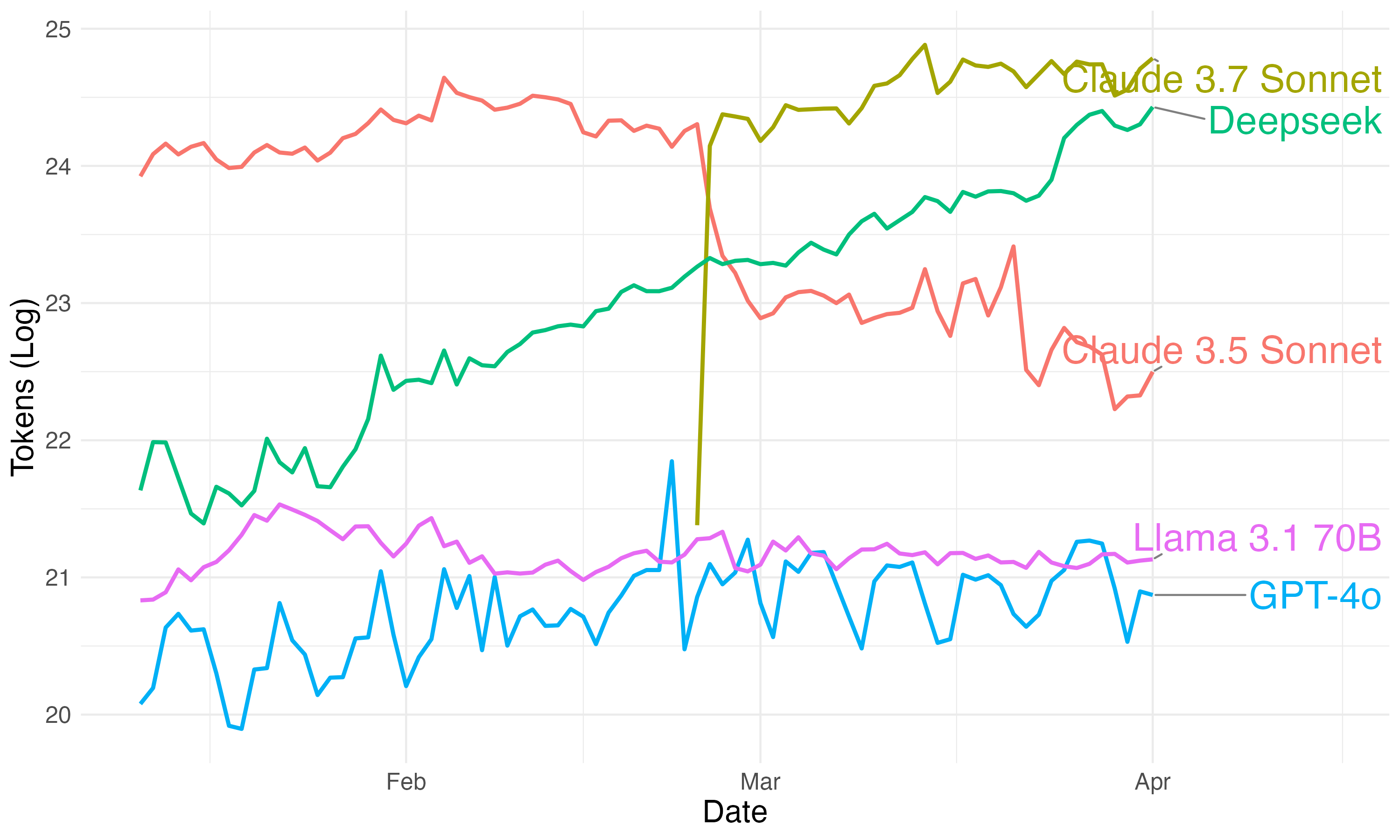}
    \caption{Claude 3.7 Sonnet Release Comparison (Log Scale)}
    \label{fig:claude_case_study_log}
\end{figure}

\begin{figure}[H]
    \centering
    \includegraphics[width=0.8\textwidth]{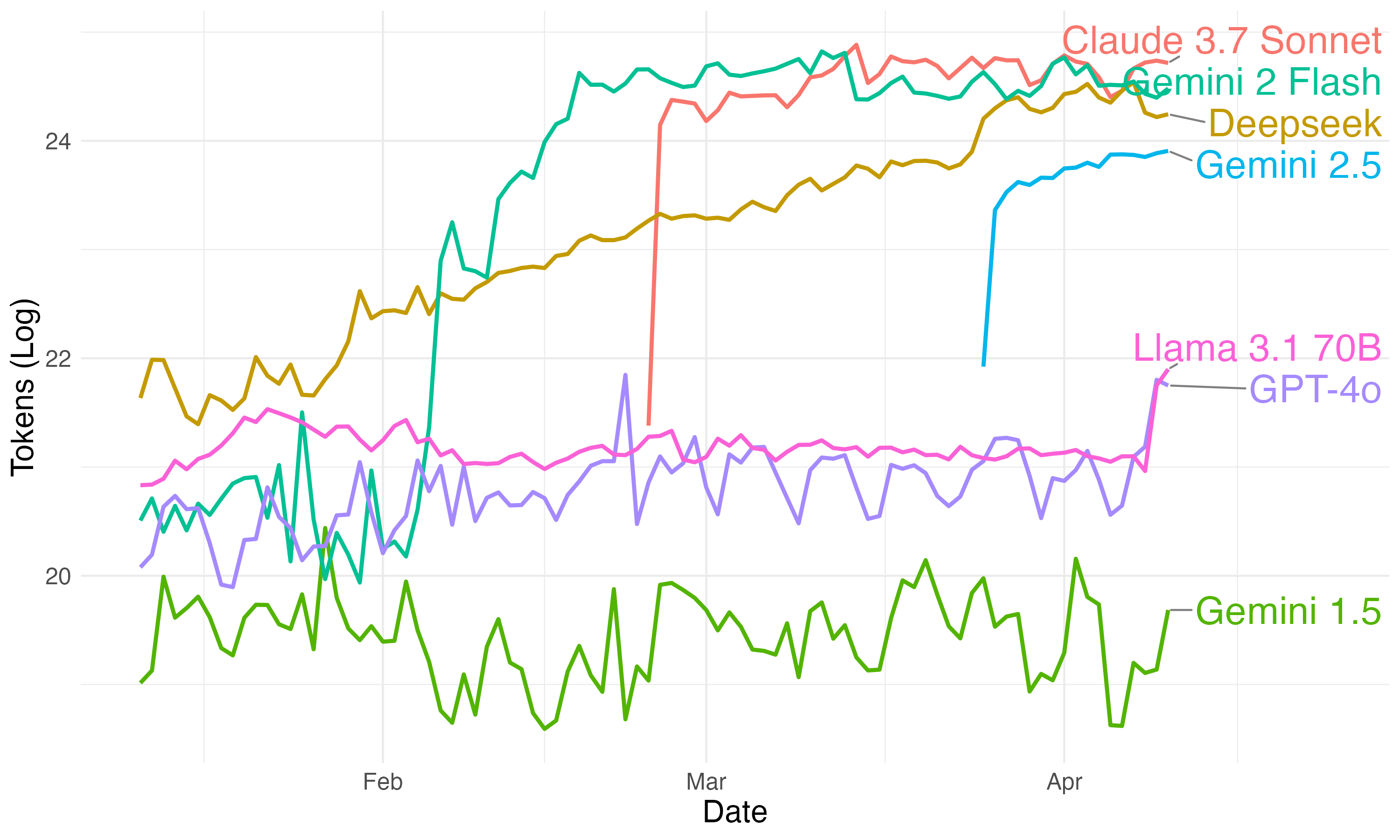}
    \caption{Gemini 2.5 Pro Release Comparison (Log Scale)}
    \label{fig:gemini25_case_study_log}
\end{figure}

\begin{figure}[H]
    \centering
    \includegraphics[width=0.8\textwidth]{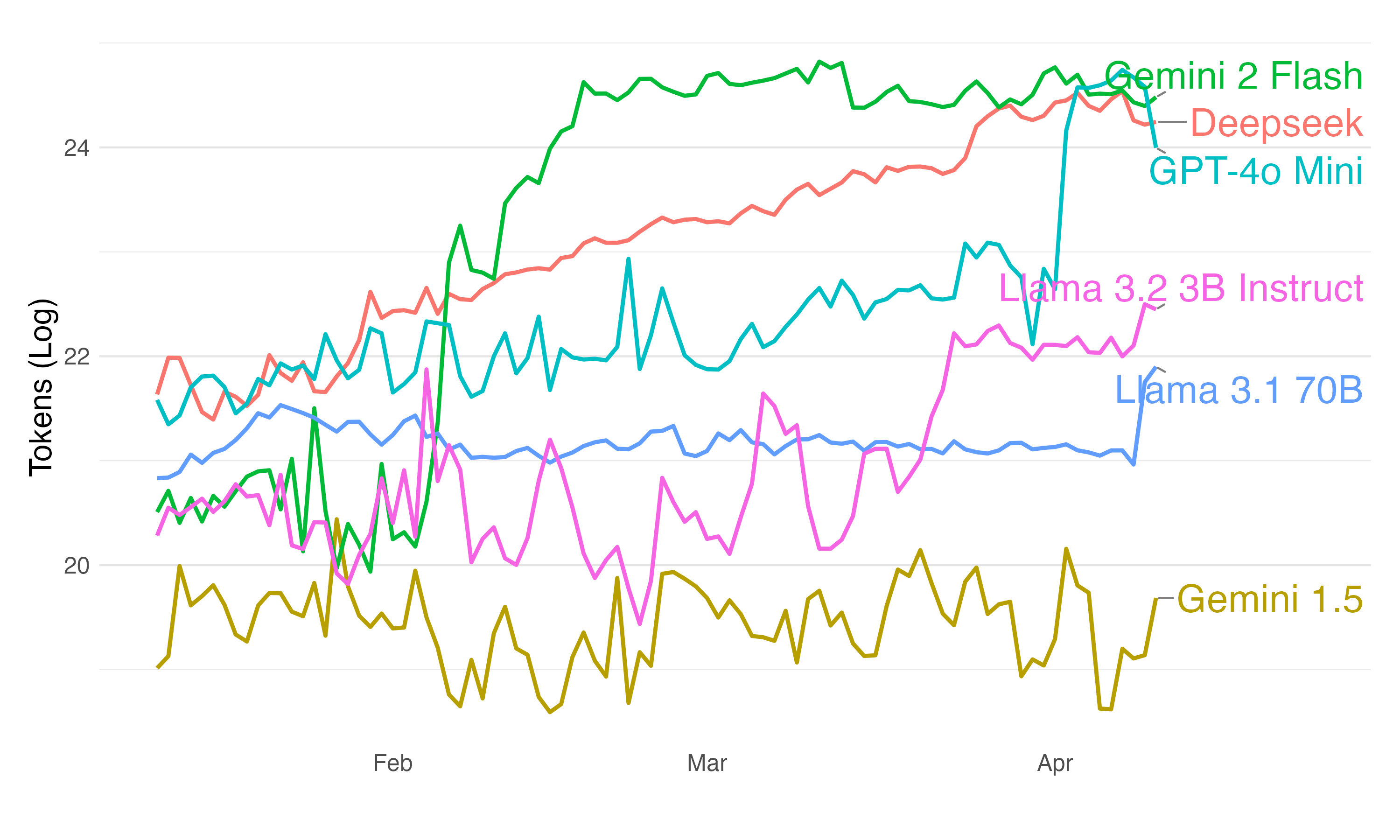}
    \caption{Gemini 2.0 Flash Release Comparison (Log Scale)}
    \label{fig:gemini_flash_case_study_log}
\end{figure}

\begin{figure}[H]
    \centering
    \includegraphics[width=0.8\textwidth]{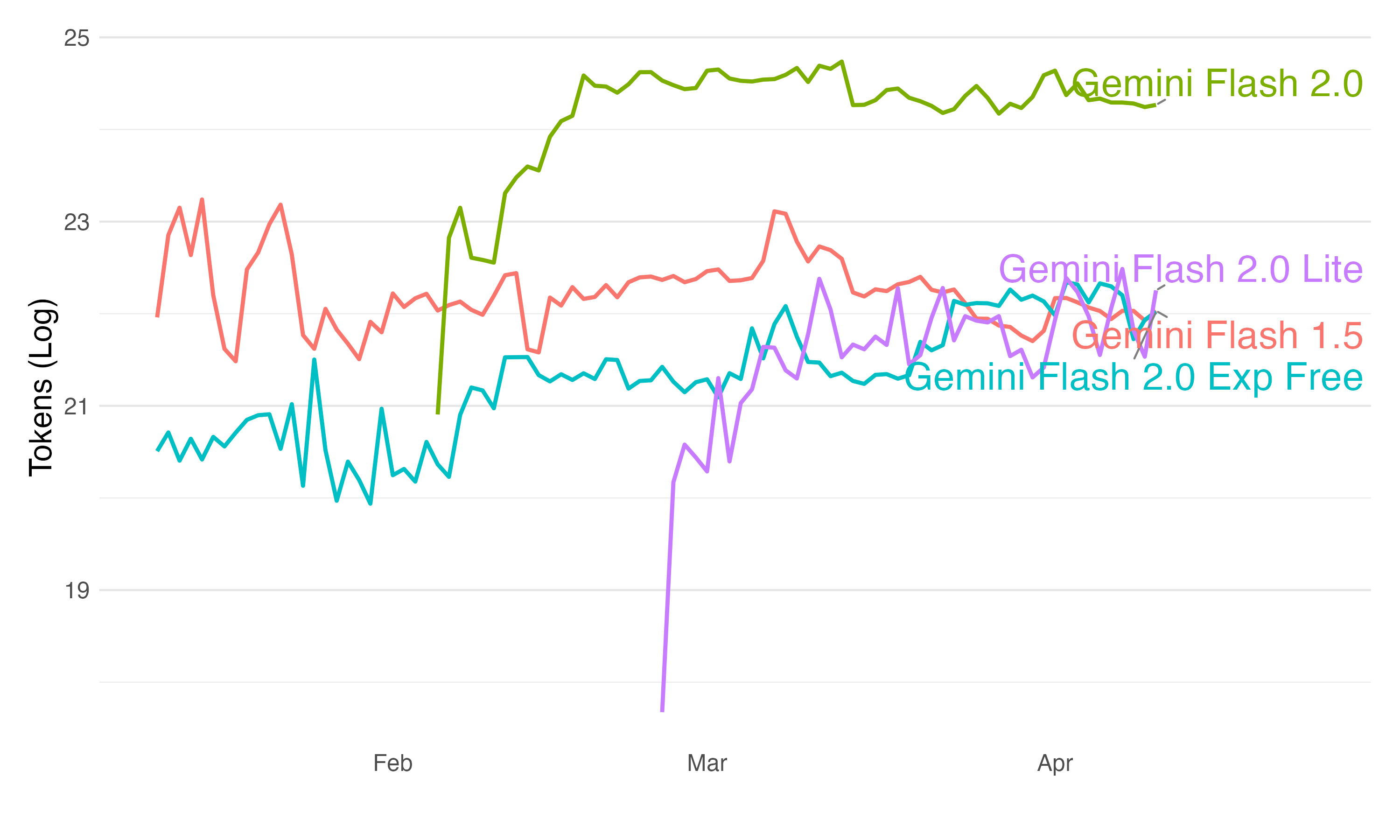}
    \caption{All Gemini 2.0 Models Usage (Log Scale)}
    \label{fig:gemini_only_log}
\end{figure}

\section{Additional Tables}

\begin{table}[htbp] 
    \centering 
    \caption{Summary Statistics by Provider} 
    \label{tab:summary_statsprovider} 
    \input{tables/summary_stats_by_provider.tex}
    
    \smallskip
    \small{\emph{Notes:} Each observation is a model by day. MTok stands for million tokens.}
\end{table}

\begin{table}[htbp] 
    \centering 
    \caption{Summary Statistics by Class} 
    \label{tab:summary_statsclass} 
    \input{tables/summary_stats_by_class.tex}
    
    \smallskip
    \small{\emph{Notes:} Each observation is a model by day. MTok stands for million tokens.}
\end{table}
\clearpage
\captionof{table}{List of All Models Used in Analysis}
\label{tab:allmodels}
\smalllongtable     
\input{tables/unique_models.tex}

\end{document}

%% file: tables/summary_stats_overall.tex
\begin{tabular}{lrrrrrr}
\toprule
& Mean & Median & SD & Min & Max & N \\ \hline
Completion Tokens (M) & \num{55.24} & \num{2.61} & \num{281.17} & \num{0.00} & \num{5069.85} & 16236 \\
Prompt Tokens (M) & \num{733.75} & \num{26.91} & \num{3911.66} & \num{0.00} & \num{55130.95} & 16236 \\
Context Window (M) & \num{0.15} & \num{0.13} & \num{0.26} & \num{0.00} & \num{2.00} & 16236 \\
Output Price per MTok & \num{5.89} & \num{0.80} & \num{25.44} & \num{0.00} & \num{600.00} & 16236 \\
Input Price per MTok & \num{1.80} & \num{0.50} & \num{7.10} & \num{0.00} & \num{150.00} & 16236 \\
\hline
\end{tabular}

%% file: tables/top_10_apps_simple.tex
\begin{tabular}{lr}
\toprule
Application & Usage (Millions of Tokens) \\
\midrule
Cline & 314583 \\
Roo Code & 231702 \\
shapes inc & 37620 \\
SillyTavern & 36883 \\
Chub AI & 17435 \\
DocsLoop & 14264 \\
OpenRouter: Chatroom & 13405 \\
liteLLM & 11461 \\
Fraction AI & 8085 \\
Fish Audio & 6445 \\
\bottomrule
\end{tabular}

%% file: tables/summary_stats_by_provider.tex
\begin{tabular}{lrrrrrrrrrrrrr}
\toprule
Provider &   & Mean & Median & SD \\ \hline
Anthropic & Completion Tokens (M) & \num{68.94} & \num{3.28} & \num{170.16} \\
& Prompt Tokens (M) & \num{3885.71} & \num{69.86} & \num{10448.85} \\
& Context Window (M) & \num{0.20} & \num{0.20} & \num{0.00} \\
& Output Price per MTok & \num{19.04} & \num{15.00} & \num{25.60} \\
& Input Price per MTok & \num{3.81} & \num{3.00} & \num{5.12} \\
Google & Completion Tokens (M) & \num{288.80} & \num{12.34} & \num{801.77} \\
& Prompt Tokens (M) & \num{2845.18} & \num{156.36} & \num{7833.41} \\
& Context Window (M) & \num{0.69} & \num{1.00} & \num{0.59} \\
& Output Price per MTok & \num{0.53} & \num{0.00} & \num{1.44} \\
& Input Price per MTok & \num{0.17} & \num{0.00} & \num{0.37} \\
OpenAI & Completion Tokens (M) & \num{48.68} & \num{4.79} & \num{168.81} \\
& Prompt Tokens (M) & \num{518.27} & \num{24.42} & \num{3317.77} \\
& Context Window (M) & \num{0.13} & \num{0.13} & \num{0.04} \\
& Output Price per MTok & \num{29.33} & \num{10.00} & \num{70.11} \\
& Input Price per MTok & \num{8.79} & \num{2.50} & \num{19.94} \\
Other & Completion Tokens (M) & \num{30.30} & \num{2.10} & \num{157.13} \\
& Prompt Tokens (M) & \num{261.11} & \num{19.25} & \num{1065.70} \\
& Context Window (M) & \num{0.10} & \num{0.07} & \num{0.13} \\
& Output Price per MTok & \num{2.04} & \num{0.60} & \num{3.47} \\
& Input Price per MTok & \num{0.83} & \num{0.25} & \num{1.12} \\
\hline
\end{tabular}

%% file: tables/summary_stats_by_class.tex
\begin{tabular}{lrrrrrrrrrrrr}
\toprule
Model Class &   & Mean & Median & Max \\ \hline
SOTA & Completion Tokens (M) & \num{125.70} & \num{26.34} & \num{4418.45} \\
& Prompt Tokens (M) & \num{1918.66} & \num{215.68} & \num{55130.95} \\
& Context Window (M) & \num{0.22} & \num{0.13} & \num{2.00} \\
& Output Price per MTok & \num{18.17} & \num{0.90} & \num{600.00} \\
& Input Price per MTok & \num{5.33} & \num{0.50} & \num{150.00} \\
Fast \& Cheap & Completion Tokens (M) & \num{89.35} & \num{3.50} & \num{5069.85} \\
& Prompt Tokens (M) & \num{856.20} & \num{44.87} & \num{53810.14} \\
& Context Window (M) & \num{0.21} & \num{0.13} & \num{1.05} \\
& Output Price per MTok & \num{0.64} & \num{0.15} & \num{4.40} \\
& Input Price per MTok & \num{0.20} & \num{0.09} & \num{1.10} \\
Old & Completion Tokens (M) & \num{36.67} & \num{1.71} & \num{623.47} \\
& Prompt Tokens (M) & \num{2412.09} & \num{29.22} & \num{49803.12} \\
& Context Window (M) & \num{0.17} & \num{0.20} & \num{0.20} \\
& Output Price per MTok & \num{31.11} & \num{15.00} & \num{75.00} \\
& Input Price per MTok & \num{7.39} & \num{5.00} & \num{15.00} \\
\hline
\end{tabular}

%% file: tables/unique_models.tex
\begin{longtable}{lll}
\toprule
Model Name & Provider & Model Class\\
\midrule
\endfirsthead
\multicolumn{3}{@{}l}{\textit{(continued)}}\\
\toprule
Model Name & Provider & Model Class\\
\midrule
\endhead

\endfoot
\bottomrule
\endlastfoot
01 ai yi large & 01-Ai & Other\\
Aetherwiing mn starcannon 12b & Aetherwiing & Other\\
Ai21 jamba 1 5 Mini & Ai21 & Fast \& Cheap\\
Ai21 jamba 1 5 large & Ai21 & Other\\
Ai21 jamba 1.6 Mini & Ai21 & Fast \& Cheap\\
\addlinespace
Ai21 jamba 1.6 large & Ai21 & Other\\
Ai21 jamba Instruct & Ai21 & Other\\
Aion 1.0 & Aion-Labs & Other\\
Aion 1.0 Mini & Aion-Labs & Fast \& Cheap\\
Llama rp llama & Aion-Labs & Fast \& Cheap\\
\addlinespace
Openhands lm 32b v0.1 & All-Hands & Other\\
Molmo 7b d Free & Allenai & Fast \& Cheap\\
Olmo 2 0325 32b Instruct & Allenai & Other\\
Magnum 72b & Alpindale & Other\\
Amazon nova Pro v1 & Amazon & Other\\
\addlinespace
Amazon nova lite v1 & Amazon & Other\\
Amazon nova micro v1 & Amazon & Other\\
Anthracite org magnum v2 72b & Anthracite-Org & Other\\
Anthracite org magnum v4 72b & Anthracite-Org & Other\\
Claude 3 Haiku & Anthropic & Fast \& Cheap\\
\addlinespace
Claude 3 Haiku beta & Anthropic & Fast \& Cheap\\
Claude 3 Opus & Anthropic & Old\\
Claude 3 Opus beta & Anthropic & Old\\
Claude 3 Sonnet & Anthropic & Old\\
Claude 3 Sonnet beta & Anthropic & Old\\
\addlinespace
Claude Haiku 3.5 & Anthropic & Fast \& Cheap\\
Claude Sonnet 3.5 & Anthropic & Old\\
Claude Sonnet 3.7 & Anthropic & SOTA\\
Claude Sonnet 3.7 Thinking & Anthropic & SOTA\\
Ui tars 72b Free & Bytedance-Research & Other\\
\addlinespace
Dolphin mixtral 8x22b & Cognitivecomputations & Other\\
Dolphin3.0 mistral 24b Free & Cognitivecomputations & Other\\
Dolphin3.0 r1 mistral 24b Free & Cognitivecomputations & Other\\
Command & Cohere & Other\\
Command a & Cohere & Other\\
\addlinespace
Command r & Cohere & Other\\
Command r 03 2024 & Cohere & Other\\
Command r 08 2024 & Cohere & Other\\
Command r Plus & Cohere & Other\\
Command r Plus 04 2024 & Cohere & Other\\
\addlinespace
Command r Plus 08 2024 & Cohere & Other\\
Command r7b 12 2024 & Cohere & Fast \& Cheap\\
Deepseek Chat & Deepseek & SOTA\\
Deepseek Chat Free & Deepseek & SOTA\\
Deepseek Chat v3 0324 & Deepseek & SOTA\\
\addlinespace
Deepseek Chat v3 0324 Free & Deepseek & SOTA\\
Deepseek r1 & Deepseek & SOTA\\
Deepseek r1 Free & Deepseek & SOTA\\
Deepseek r1 distill llama 70b & Deepseek & SOTA\\
Deepseek r1 distill llama 70b Free & Deepseek & SOTA\\
\addlinespace
Deepseek r1 distill llama 8b & Deepseek & Fast \& Cheap\\
Deepseek r1 distill qwen 1.5b & Deepseek & Fast \& Cheap\\
Deepseek r1 distill qwen 14b & Deepseek & Fast \& Cheap\\
Deepseek r1 distill qwen 14b Free & Deepseek & Fast \& Cheap\\
Deepseek r1 distill qwen 32b & Deepseek & SOTA\\
\addlinespace
Deepseek r1 distill qwen 32b Free & Deepseek & SOTA\\
Deepseek r1 zero Free & Deepseek & SOTA\\
Deepseek v3 base Free & Deepseek & SOTA\\
Eva qwen 2.5 32b & Eva-Unit-01 & Other\\
Eva qwen 2.5 72b & Eva-Unit-01 & Other\\
\addlinespace
Llama llama 3.33 & Eva-Unit-01 & Other\\
Qwerky 72b Free & Featherless & Other\\
Gemini Flash 1.5 & Google & Fast \& Cheap\\
Gemini Flash 1.5 8B & Google & Fast \& Cheap\\
Gemini Flash 1.5 Exp 8B & Google & Fast \& Cheap\\
\addlinespace
Gemini Flash 2.0 & Google & Fast \& Cheap\\
Gemini Flash 2.0 Exp Free & Google & Fast \& Cheap\\
Gemini Flash 2.0 Lite & Google & Fast \& Cheap\\
Gemini Flash 2.0 Thinking Free & Google & Fast \& Cheap\\
Gemini Pro 1.5 & Google & SOTA\\
\addlinespace
Gemini Pro 2.5 Exp Free & Google & SOTA\\
Gemini Pro 2.5 Preview & Google & SOTA\\
Gemma 2 27b it & Google & Fast \& Cheap\\
Gemma 2 9b it & Google & Fast \& Cheap\\
Gemma 2 9b it Free & Google & Fast \& Cheap\\
\addlinespace
Gemma 3 12b it & Google & Fast \& Cheap\\
Gemma 3 12b it Free & Google & Fast \& Cheap\\
Gemma 3 1b it Free & Google & Fast \& Cheap\\
Gemma 3 27b it & Google & Fast \& Cheap\\
Gemma 3 27b it Free & Google & Fast \& Cheap\\
\addlinespace
Gemma 3 4b it & Google & Fast \& Cheap\\
Gemma 3 4b it Free & Google & Fast \& Cheap\\
Learnlm 1.5 Pro experimental Free & Google & Other\\
Infermatic mn inferor 12b & Infermatic & Other\\
Inflection 3 pi & Inflection & Other\\
\addlinespace
Inflection 3 productivity & Inflection & Other\\
Llama large 70b & Latitudegames & Other\\
Lfm 3b & Liquid & Fast \& Cheap\\
Lfm 40b & Liquid & Other\\
Lfm 7b & Liquid & Fast \& Cheap\\
\addlinespace
Llama 3 70b Instruct & Meta-Llama & SOTA\\
Llama 3 8b Instruct & Meta-Llama & Fast \& Cheap\\
Llama 3.1 405b & Meta-Llama & Other\\
Llama 3.1 70b & Meta-Llama & SOTA\\
Llama 3.1 8b & Meta-Llama & Fast \& Cheap\\
\addlinespace
Llama 3.2 11b & Meta-Llama & Other\\
Llama 3.2 1b & Meta-Llama & Other\\
Llama 3.2 3b & Meta-Llama & Fast \& Cheap\\
Llama 3.2 90b & Meta-Llama & Other\\
Llama 3.3 70b & Meta-Llama & SOTA\\
\addlinespace
Llama 4 maverick & Meta-Llama & Other\\
Llama 4 maverick Free & Meta-Llama & Other\\
Llama 4 scout & Meta-Llama & Other\\
Llama 4 scout Free & Meta-Llama & Other\\
Llama guard 2 8b & Meta-Llama & Fast \& Cheap\\
\addlinespace
Llama guard 3 8b & Meta-Llama & Fast \& Cheap\\
Phi 3 Mini 128k Instruct & Microsoft & Fast \& Cheap\\
Phi 3 medium 128k Instruct & Microsoft & Other\\
Phi 3.5 Mini 128k Instruct & Microsoft & Fast \& Cheap\\
Phi 4 & Microsoft & Other\\
\addlinespace
Phi 4 multimodal Instruct & Microsoft & Other\\
Wizardlm 2 7b & Microsoft & Fast \& Cheap\\
Wizardlm 2 8x22b & Microsoft & SOTA\\
Minimax 01 & Minimax & Fast \& Cheap\\
Mistral ministral 8b & Mistral & Fast \& Cheap\\
\addlinespace
Codestral 2501 & Mistralai & SOTA\\
Codestral Mamba & Mistralai & SOTA\\
Ministral 3b & Mistralai & Fast \& Cheap\\
Ministral 8b & Mistralai & Fast \& Cheap\\
Mistral large & Mistralai & Other\\
\addlinespace
Mistral large 2407 & Mistralai & Other\\
Mistral large 2411 & Mistralai & Other\\
Mistral medium & Mistralai & Other\\
Mistral nemo & Mistralai & Fast \& Cheap\\
Mistral nemo Free & Mistralai & Fast \& Cheap\\
\addlinespace
Mistral saba & Mistralai & Other\\
Mistral small & Mistralai & Fast \& Cheap\\
Mistral small 24b Instruct 2501 & Mistralai & Fast \& Cheap\\
Mistral small 24b Instruct 2501 Free & Mistralai & Fast \& Cheap\\
Mistral small 3.1 24b Instruct & Mistralai & Fast \& Cheap\\
\addlinespace
Mistral small 3.1 24b Instruct Free & Mistralai & Fast \& Cheap\\
Mistral tiny & Mistralai & Fast \& Cheap\\
Mixtral 8x22b Instruct & Mistralai & Other\\
Pixtral 12b & Mistralai & Other\\
Pixtral large 2411 & Mistralai & Other\\
\addlinespace
Kimi vl a3b thinking Free & Moonshotai & Fast \& Cheap\\
Moonlight 16b a3b Instruct Free & Moonshotai & Fast \& Cheap\\
Llama 3 lumimaid 70b & Neversleep & Other\\
Llama 3 lumimaid 8b & Neversleep & Fast \& Cheap\\
Llama 3 lumimaid 8b extended & Neversleep & Fast \& Cheap\\
\addlinespace
Llama 3.1 lumimaid & Neversleep & Other\\
Llama 3.1 lumimaid & Neversleep & Fast \& Cheap\\
Nothingiisreal mn celeste 12b & Nothingiisreal & Other\\
Deephermes 3 llama 3 8b Preview Free & Nousresearch & Fast \& Cheap\\
Hermes 2 Pro llama 3 8b & Nousresearch & Fast \& Cheap\\
\addlinespace
Llama 3 llama & Nousresearch & Other\\
Llama 3 llama & Nousresearch & SOTA\\
Nous hermes 2 mixtral 8x7b dpo & Nousresearch & Fast \& Cheap\\
Llama llama 3.1 & Nvidia & Fast \& Cheap\\
Llama llama 3.3 & Nvidia & Fast \& Cheap\\
\addlinespace
Olympiccoder 32b Free & Open-R1 & Other\\
Olympiccoder 7b Free & Open-R1 & Fast \& Cheap\\
ChatGPT-4o latest & Openai & SOTA\\
GPT-3.5 Turbo 0613 & Openai & Other\\
GPT-4 Turbo & Openai & Old\\
\addlinespace
GPT-4 Turbo Preview & Openai & Old\\
GPT-4.5 Preview & Openai & SOTA\\
GPT-4o & Openai & SOTA\\
GPT-4o 2024 05 13 & Openai & Old\\
GPT-4o 2024 08 06 & Openai & Old\\
\addlinespace
GPT-4o 2024 11 20 & Openai & SOTA\\
GPT-4o Mini & Openai & Fast \& Cheap\\
GPT-4o Mini 2024 07 18 & Openai & Fast \& Cheap\\
GPT-4o Mini search Preview & Openai & Fast \& Cheap\\
GPT-4o extended & Openai & SOTA\\
\addlinespace
GPT-4o search Preview & Openai & SOTA\\
O1 & Openai & SOTA\\
O1 Mini & Openai & Fast \& Cheap\\
O1 Mini 2024 09 12 & Openai & Fast \& Cheap\\
O1 Preview & Openai & SOTA\\
\addlinespace
O1 Preview 2024 09 12 & Openai & SOTA\\
O1 Pro & Openai & SOTA\\
O3 Mini & Openai & SOTA\\
O3 Mini high & Openai & SOTA\\
Optimus alpha & Openrouter & Other\\
\addlinespace
Llama 3.1 sonar & Perplexity & Other\\
R1 1776 & Perplexity & Other\\
Sonar & Perplexity & Other\\
Sonar Pro & Perplexity & Other\\
Sonar deep research & Perplexity & Other\\
\addlinespace
Sonar reasoning & Perplexity & Other\\
Sonar reasoning Pro & Perplexity & Other\\
Qwen 2 72b Instruct & Qwen & Other\\
Qwen 2.5 72b Instruct & Qwen & Other\\
Qwen 2.5 72b Instruct Free & Qwen & Other\\
\addlinespace
Qwen 2.5 7b Instruct & Qwen & Fast \& Cheap\\
Qwen 2.5 7b Instruct Free & Qwen & Fast \& Cheap\\
Qwen 2.5 coder 32b Instruct & Qwen & Other\\
Qwen 2.5 coder 32b Instruct Free & Qwen & Other\\
Qwen 2.5 vl 72b Instruct & Qwen & Other\\
\addlinespace
Qwen 2.5 vl 7b Instruct & Qwen & Fast \& Cheap\\
Qwen 2.5 vl 7b Instruct Free & Qwen & Fast \& Cheap\\
Qwen Plus & Qwen & Other\\
Qwen Turbo & Qwen & Other\\
Qwen max & Qwen & Other\\
\addlinespace
Qwen vl Plus & Qwen & Other\\
Qwen vl max & Qwen & Other\\
Qwen2.5 32b Instruct & Qwen & Other\\
Qwen2.5 vl 32b Instruct & Qwen & Other\\
Qwen2.5 vl 32b Instruct Free & Qwen & Other\\
\addlinespace
Qwen2.5 vl 3b Instruct Free & Qwen & Fast \& Cheap\\
Qwen2.5 vl 72b Instruct & Qwen & Other\\
Qwen2.5 vl 72b Instruct Free & Qwen & Other\\
Qwq 32b & Qwen & Other\\
Qwq 32b Free & Qwen & Other\\
\addlinespace
Qwq 32b Preview & Qwen & Other\\
Qwq 32b Preview Free & Qwen & Other\\
Sorcererlm 8x22b & Raifle & Other\\
Reka Flash 3 Free & Rekaai & Fast \& Cheap\\
Fimbulvetr 11b v2 & Sao10k & Other\\
\addlinespace
L3 euryale 70b & Sao10k & Other\\
L3 lunaris 8b & Sao10k & Fast \& Cheap\\
L3.1 70b hanami x1 & Sao10k & Other\\
L3.1 euryale 70b & Sao10k & Other\\
L3.3 euryale 70b & Sao10k & Other\\
\addlinespace
Llama3.1 typhoon2 70b Instruct & Scb10x & Other\\
Llama3.1 typhoon2 8b Instruct & Scb10x & Fast \& Cheap\\
Midnight rose 70b & Sophosympatheia & Other\\
Rogue rose 103b v0.2 Free & Sophosympatheia & Fast \& Cheap\\
L3.3 electra r1 70b & Steelskull & Other\\
\addlinespace
Anubis Pro 105b v1 & Thedrummer & Fast \& Cheap\\
Rocinante 12b & Thedrummer & Other\\
Skyfall 36b v2 & Thedrummer & Other\\
Unslopnemo 12b & Thedrummer & Fast \& Cheap\\
Llama 3.1 swallow & Tokyotech-Llm & Other\\
\addlinespace
Llama 3.1 swallow & Tokyotech-Llm & Fast \& Cheap\\
Grok 2 1212 & X-Ai & Other\\
Grok 2 Vision 1212 & X-Ai & Other\\
Grok 3 Mini beta & X-Ai & Fast \& Cheap\\
Grok 3 beta & X-Ai & SOTA\\
\addlinespace
Grok Vision beta & X-Ai & Other\\
Grok beta & X-Ai & Other\\*
\end{longtable}